\documentclass[aps,prc,onecolumn,superscriptaddress,showpacs,showkeys]{revtex4}
\usepackage{graphicx}
\usepackage{bm}
\usepackage{epsfig}
\usepackage{amsfonts}
\usepackage{amssymb,amscd}
\usepackage{subfigure}

\begin{document}

\title{On the rapidity dependence of the average transverse momentum in hadronic collisions}

\author{F.O. Dur\~aes}
\affiliation{Curso de F\'{\i}sica, Escola de Engenharia, Universidade Presbiteriana Mackenzie\\
CEP 01302-907, S\~{a}o Paulo, Brazil}
\author{A.V. Giannini}
\affiliation{Instituto de F\'{\i}sica, Universidade de S\~{a}o Paulo\\
C.P. 66318,  05315-970 S\~{a}o Paulo, SP, Brazil}
\author{V.P.  Gon\c{c}alves}
\affiliation{Instituto de F\'{\i}sica e Matem\'atica,  Universidade Federal de Pelotas\\
Caixa Postal 354, CEP 96010-900, Pelotas, RS, Brazil}
\affiliation{Department of Astronomy and Theoretical Physics, Lund University, 223-62 Lund, Sweden.}
\author{F. S. Navarra}
\affiliation{Instituto de F\'{\i}sica, Universidade de S\~{a}o Paulo\\
C.P. 66318,  05315-970 S\~{a}o Paulo, SP, Brazil}

\begin{abstract}
The energy and rapidity dependence of the average transverse momentum $\langle p_T \rangle$ in $pp$ and $pA$ collisions at RHIC and 
LHC energies are estimated using the Colour Glass Condensate (CGC) formalism. 
We update previous predictions for the $p_T$ - spectra using the hybrid formalism of the CGC approach 
and two phenomenological models for the dipole - target scattering amplitude. We demonstrate that these 
models are able to  describe the RHIC and LHC data  for the hadron production in $pp$, $dAu$ and $pPb$ 
collisions at $p_T \le 20$ GeV. Moreover, we present our predictions for  $\langle p_T \rangle$ and 
demonstrate that the ratio $\langle p_{T}(y)\rangle / \langle p_{T}(y = 0)\rangle$  decreases with the 
rapidity and has a  behaviour similar to that predicted  by hydrodynamical calculations. 

\end{abstract}
\maketitle
\vspace{1cm}

\section{Introduction}

\date{\today}

The Large Hadron Collider (LHC) has opened up a new frontier in high energy hadron - hadron collisions, 
allowing to test the Quantum Chromodynamics in unexplored regimes of energy, density and rapidities, 
considering different configurations of the colliding hadrons (protons and nuclei) (For a recent review 
see e.g. \cite{hdqcd}). In particular, the LHC experiments  have unprecedented capacities to study 
several subjects associated to {\it forward physics}  as, for instance, soft and hard diffraction, 
exclusive production of new mass states, low-$x$ dynamics and other important topics (For a review see 
e.g. Ref. \cite{blois}). Forward physics is characterized by the production of particles with relatively 
small transverse momentum, being traditionally associated with soft particle production, which is intrinsically 
non perturbative and not amenable to first-principles analysis. However, in the particle production at large 
energies and forward rapidities, the wave function of one of the projectiles is probed at large Bjorken $x$ 
and that of the other at very small $x$. The latter  is characterized by a large number of gluons, which is 
expected to form a new state of matter - the Colour Glass Condensate (CGC) -  where the gluon distribution 
saturates and non linear coherence phenomena dominate \cite{hdqcd}. Such a system is endowed with a new dynamical 
momentum scale, the saturation scale $Q_s$, which controls the main features of  particle production 
and whose evolution is described by an infinite hierarchy of coupled equations for the correlators of Wilson 
lines \cite{BAL,KOVCHEGOV,CGC}.  At  large energies and rapidities, $Q_s$ is expected to become much larger 
than the QCD confinement scale $\Lambda_{QCD}$. Furthermore, the saturation scale is expected to determine the 
typical transverse momentum of the  produced partons in the interaction. Consequently, the probe of the average 
transverse momentum $\langle p_T \rangle$ in hadronic collisions can provide important information about the QCD 
dynamics (For related studies see, e.g. Refs. \cite{jamal,amirpt,mclpt}).

Another motivation for a detailed analysis of $\langle p_T \rangle$ in $pp$ and $pA$ collisions 
is the recent suggestion made in Ref. \cite{bbs} that this quantity can be used to disentangle 
the hydrodynamic and the CGC descriptions of  the  ``ridge'' effect (the appearance of long range 
correlations in the relative pseudorapidity $\Delta\eta$ and the relative azimuthal angle $\Delta\phi$ 
plane) observed  in high multiplicity events in small colliding systems 
such as $pp$ and $p(d)A$ \cite{cms,cms2,alice1,atlas,alice2,phenix,star,cms3}. While the 
previously ridge-type structure observed in heavy-ion collisions at RHIC and the LHC 
was considered as an evidence of the hydrodynamical nature of the quark-gluon-plasma, 
see e.g. Refs. \cite{alver,star2} there is no compelling reason why 
small systems should also exhibit a hydrodynamical behaviour 
even though a hydro approach is able to describe the 
experimental data \cite{broni,broni2}. On the other hand, the CGC approach also provides a qualitatively 
good description of the same data \cite{dumi,dumi2,dumi3,kov,lev,iancu,dus,yuri,ray,dumi4,reza}. 
Therefore, the origin of the ridge in $pp$ and $pA$ collisions is still an open question.
As the ridge effect, the azimuthal asymmetries observed in pPb collisions 
at the LHC energies by the ALICE \cite{alice1}, ATLAS \cite{atlas,ATLAS_vn}
and CMS \cite{cms2,cms3} collaborations are also open to different theoretical 
explanations. While in the hydro approaches those anisotropies emerge as a final state 
feature due to the hydrodynamic flow \cite{broni,broni2,hydro_vn} in the 
CGC approach they are described as a initial state anisotropies which are present 
at the earliest stages of the collision \cite{CGC_vn}.
In Ref. \cite{bbs}, the authors have studied the rapidity ($y$) dependence of the average transverse momentum of charged 
particles using very general arguments that lead to simple analytical expressions. In particular,  
the Golec - Biernat -- Wusthoff (GBW) model \cite{gbw} was used to describe the unintegrated gluon distribution and  the 
fragmentation of the partons into  final state particles was neglected. The authors of \cite{bbs} have found that the 
average transverse momentum $\langle p_T \rangle$  in the CGC approach grows with  rapidity, 
in contrast to what is expected from a collective expansion. Indeed, the hydrodynamical 
model predicts a decrease of the average transverse momentum when going from midrapidity, $y = 0$, 
to the proton side, owing to a decreasing number of produced particles. The prediction of these 
distinct behaviours is one the main motivations for the detailed analysis of the energy and rapidity 
dependencies of $\langle p_T \rangle$ and thus to verify  how robust this conclusion is. 
As the GBW model does not describe the $p_T$ - spectra measured in $pp/pA$ collisions, in our study we will consider two more 
realistic phenomenological saturation models that  are able to reproduce the experimental data in the region of small transverse 
momenta. This is the region that determines the behavior of the average transverse momentum. Moreover, we will analyse the impact 
of the inclusion of  parton fragmentation in the rapidity dependence of $\langle p_T \rangle$. With these improvements, we are able to  
present realistic  predictions for  $\langle p_T \rangle$ based on the CGC results that 
are able to describe the current experimental data on hadron production in hadronic collisions.

This paper is organized as follows. In the next Section we present a brief review of the hybrid 
formalism and discuss the phenomenological models of the dipole scattering amplitudes used in our analysis. 
In Section \ref{section:results} we update the main parameters of these phenomenological models by the 
comparison with the RHIC and LHC data on hadron production in $pp$, $dAu$ and $pPb$ collisions. Using 
the new version of these  models, which are able to describe the experimental data for $p_T \le 20$ GeV, 
we present our predictions for the rapidity and energy dependencies of the average transverse momentum in 
$pp$ and $pPb$ collisions.  Finally, in Section \ref{section:conc} we summarize our main conclusions.

\section{Particle  production in the CGC: the hybrid Formalism}
\label{section:formulario}

In order to estimate the energy and rapidity dependencies of the average transverse momentum $\langle p_T \rangle$ 
we will need to describe  particle production at forward rapidities and large energies. 
The description of  hadron production at large transverse momentum $p_T$ is one the main examples of a hard process in perturbative 
QCD (pQCD). It  can be accurately described within collinear factorization, by combining partonic cross-sections computed to some 
fixed order in perturbation theory with parton distribution and fragmentation functions whose evolution is computed by solving the 
Dokshitzer - Gribov - Lipatov - Altarelli - Parisi (DGLAP) equations \cite{dglap} to the corresponding accuracy in pQCD.  The high 
transverse momentum $p_T$ of the produced hadron ensures the applicability of pQCD, which is expected to fail to low-$p_T^2$. Furthermore, 
at forward rapidities the small-$x$ evolution becomes important, leading to a growth of the gluon density  and of the gluon  transverse 
momentum. Because of that, in this kinematical range their evolution in transverse momentum cannot be disregarded, which implies that at 
very forward rapidities the collinear factorization is expected to break down. An alternative is the description of  hadron production 
using the $k_T$-factorization scheme, which is based on the unintegrated gluon distributions whose evolution is described by  the 
Balitsky-Fadin-Kuraev-Lipatov (BFKL) equation \cite{bfkl}. However, if the transverse momentum of some of the produced particles is 
comparable with the saturation momentum scale, the partons from one projectile scatter off a dense gluonic system in the other projectile. 
In this case the  parton undergoes multiple scatterings, which cannot be encoded in the traditional (collinear and $k_T$) factorization schemes.  
As pointed in Ref. \cite{difusivo}, the forward hadron production in hadron-hadron collisions is  a typical example of a dilute-dense process,
 which is an ideal system to study the small-$x$ components of the  target wave function.  In this case the cross section is expressed as a 
convolution of the standard parton distributions for the dilute projectile, the dipole-hadron scattering amplitude (which includes the 
high-density effects) and the parton fragmentation functions.  Basically, assuming this generalized dense-dilute factorization, the minimum 
bias invariant yield for single-inclusive hadron production in hadron-hadron processes is described in the CGC formalism  by 
 \cite{dhj}
\begin{eqnarray}
{dN_h \over dy d^2p_T} &=& 
{K(y) \over (2\pi)^2} \int_{x_F}^{1} dx_1 \, {x_1\over x_F}
\Bigg[f_{q/p}(x_1,\mu^2)\, \widetilde{\cal N}_{F} \left({x_1\over x_F}p_T,x_2\right)\,
D_{h/q}\left({x_F\over x_1},\mu^2\right)
\nonumber \\
& &+~
f_{g/p}(x_1,\mu^2)\, \widetilde{\cal N}_{A} \left({x_1\over x_F}p_T,x_2\right)\, 
D_{h/g}\left({x_F\over x_1},\mu^2\right)\Bigg]~
\label{hybel}\,,
\end{eqnarray}
where $p_T$, $y$ and $x_F$ are the transverse momentum, rapidity and the Feynman-$x$
of the produced hadron, respectively. The $K(y)$-factor mimics the effect of higher order corrections and, effectively, of other dynamical 
effects not included in the CGC formulation.  The variable $x_1$ denotes the momentum
fraction of a projectile parton,    $f(x_1,\mu^2)$ the projectile parton
distribution functions  and $D(z, \mu^2)$ the parton fragmentation
functions into hadrons. These quantities  evolve according to the 
DGLAP evolution equations \cite{dglap} and obey the momentum
sum-rule. It is useful to assume $\mu^2 = p_T^2$.  Moreover, $x_F=\frac{p_T}{\sqrt{s}}e^{y}$ and the momentum fraction of the target partons 
is given by $x_2=x_1e^{-2y}$ (For details see e.g. \cite{dhj}).
 In Eq. (\ref{hybel}), $\widetilde{{\cal{N}}}_F(x,k)$  and  $\widetilde{{\cal{N}}}_A (x,k)$ are the fundamental and adjoint representations of 
the forward dipole amplitude in momentum space and are given by
\begin{eqnarray}
\widetilde{{\cal{N}}}_{A,F}(x,p_T)=  \int d^2 r \,  e^{i\vec{p_T}\cdot \vec{r}}\left[1-{\cal{N}_{A,F}}(x,r)\right]\,\,,
\end{eqnarray}
where  ${\cal{N}_{A,F}}(x,r)$  
encodes all the
information about the hadronic scattering, and thus about the
non-linear and quantum effects in the hadron wave function.
Following \cite{buw}, we will assume in what follows that ${\cal N}_{F}(x,r)$ can be  obtained from 
${\cal N}_{A}(x,r)$ after rescaling the saturation scale by $Q_{s,F}^2 = (C_{F}/C_{A}) Q_{s,A}^2 $ 
where $C_{F}/C_{A} = 4/9$. In principle, we should also include in (\ref{hybel}) the inelastic term 
that has been calculated in~\cite{tolga}. This term accounts 
for part of the full next-to-leading order correction to the hybrid formalism 
which has been recently presented in~\cite{chi-11,chi-12}. 
It has also been shown recently \cite{dumal}
that the inclusion of this term modifies  the shape of the $p_{T}$ spectra.  
However we are concerned only with the average transverse momentum $\langle p_{T} \rangle$ 
(and its  rapidity dependence) and this term plays a negligible role in this observable, 
which is dominated by the low $p_{T}$ part of the spectra. Because of this reason we will 
omit this term in our analysis.

In general the scattering amplitude ${\cal{N}_A}(x,r)$ can be obtained by solving the BK 
evolution equation \cite{BAL,KOVCHEGOV}. The BK equation is the simplest non linear 
evolution equation for the dipole-hadron scattering amplitude, being actually a mean field 
version of the first equation of the B-JIMWLK hierarchy \cite{BAL,CGC}.  
Over the last years, several authors have studied the solution of the BK equation including higher order corrections 
\cite{bkrunning,weigert,la11} and used it as input in the analysis of  leading hadron production in $pp/pA$ collisions, 
obtaining a very good description of the experimental data \cite{dumal,amir,lappi}. In what follows, instead of the solution of the BK equation, 
we will consider  two different phenomenological models based on the analytical solutions of this equation. This allows us to investigate  
 the possibility of getting a first insight on  whether or not 
the LHC data are sensitive to geometric scaling violations at high values of $p_{T}$.
Moreover, as these phenomenological models differ from the GBW model in the dependence of the anomalous dimension with 
the momentum scale (see below), it becomes possible to clarify the origin of the differences between the predictions. 
Such analysis is a hard task when the numerical solution of the BK equation is considered as input of the calculations. Our   
phenomenological approach has limitations, being valid only in a limited region of transverse momenta,  not being competitive with recent 
parametrizations that have being used to describe the nuclear modification ratio $R_{pA}$ \cite{albamar,dumal,amir,lappi,tribedy}. 
As demonstrated in those references,  a precise treatment of the nuclear geometry and/or the initial conditions is necessary  to describe the 
ratio $R_{pA}$. Such aspects are beyond the scope of this paper. On the other hand, as $\langle p_{T} \rangle$ is 
determined by the region of small $p_T$, the use of phenomenological models that describe the experimental data in this kinematical range, allows us to 
obtain realistic predictions for this quantity.
It is well known that several groups have constructed phenomenological models for the dipole 
scattering amplitude using the RHIC and/or HERA data to fix the free parameters 
\cite{dhj,buw,kkt,dips}. In general, it is  assumed that ${\cal{N}}$ can  be modelled 
through a simple Glauber-like formula,
\begin{eqnarray}
{\cal{N}}(x,r_T) = 1 - \exp\left[ -\frac{1}{4} (r_T^2 Q_s^2)^{\gamma} \right] \,\,,
\label{ngeral}
\end{eqnarray}
where $\gamma$ is the anomalous dimension of the target gluon distribution. The speed with which we move  
from the non linear regime to the extended geometric scaling regime and then from the latter to the linear
regime is what differs one phenomenological model from another. This transition speed is dictated by 
the behaviour of the anomalous dimension $\gamma (x,r_T^2)$.  In the GBW model, $\gamma$ is assumed to be constant and equal to one.  
In this paper we will consider the dipole models proposed in Refs. \cite{dhj,buw} to describe the $p_T$ 
spectra of particle production  at RHIC. In the DHJ model \cite{dhj}, the anomalous dimension was proposed to be given by
\begin{eqnarray}
\label{dip_adimension_dhj}
\gamma(x,r_T)_{DHJ}  = \gamma_{s} + (1-\gamma_s)\frac{|\log(1/r_T^{2}Q_{s}^{2})|} {\lambda y + d \sqrt{y} 
+ |\log(1/r_T^{2}Q_{s}^{2})|}.
\end{eqnarray}
with  $Q_{s}^{2} = A^{1/3} Q_{0}^{2}(x_{0}/x_{2})^{\lambda}$, $\gamma_s = 0.628$, $Q_0^2 = 1.0$ GeV$^2$, 
$x_0 = 3.0 \cdot 10^{-4}$, $\lambda = 0.288$ and $d = 1.2$. This model was designed to describe the 
forward $dAu$ data at the RHIC highest energy  taking into account geometric scaling violations 
characterized by  terms depending on the target rapidity, $y = \log(1/x_{2})$, in its 
parametrization of the anomalous dimension, with the parameter $d$ controlling the strength 
of the subleading term in $y$. In contrast, in the BUW model  \cite{buw} the anomalous 
dimension is given by
\begin{eqnarray}
\label{dip_adimension_buw}
\gamma(\omega = q_{T}/Q_{s})_{BUW} = \gamma_{s} + (1-\gamma_s)\frac{(\omega^a-1)}{(\omega^a-1)+b}\,\,,
\end{eqnarray}
where $q_{T} = p_{T}/z$ is the parton momentum. The parameters of the model ($\gamma_s = 0.628$, $a = 2.82$ and $b = 168$) 
have been fixed by fitting the $p_{T}$-spectra of the produced hadrons measured in $pp$ and $dAu$ 
collisions at the RHIC energies \cite{buw,mv08}. With these parameters the model was also able to 
describe the $ep$ HERA data for the  proton structure function if the light quark masses are neglected.   
An important feature of this model is the fact that it explicitly satisfies the property of 
geometric scaling \cite{scal,masch,prl} which is predicted by the solutions of the BK equation in 
the asymptotic regime of large energies. 
Since the forward RHIC data for the $p_T$-spectra are reproduced by both models \cite{buw,dhj}, it was 
not possible to say whether experimental data show violations of the geometric scaling or not. In principle, 
it is expected that by considering the transverse momentum distribution of produced hadrons measured 
at the LHC energies it should be possible to address this question since the new data are taken at a wider 
range of $p_T$ when compared to the RHIC data.

\begin{figure}[t]
\begin{center}
\subfigure[ ]{\label{fig:first1_alice}
\includegraphics[width=0.45\textwidth]{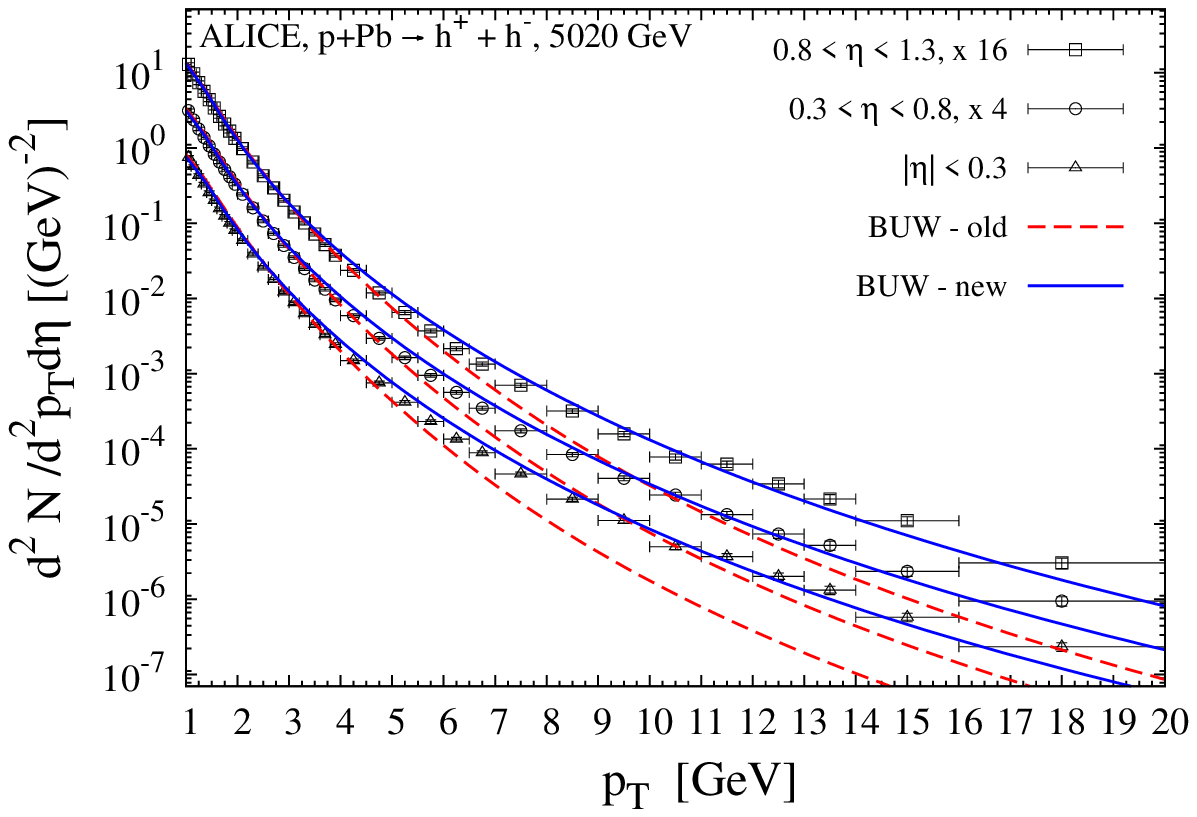}}
\subfigure[ ]{\label{fig:first2_alice}
\includegraphics[width=0.45\textwidth]{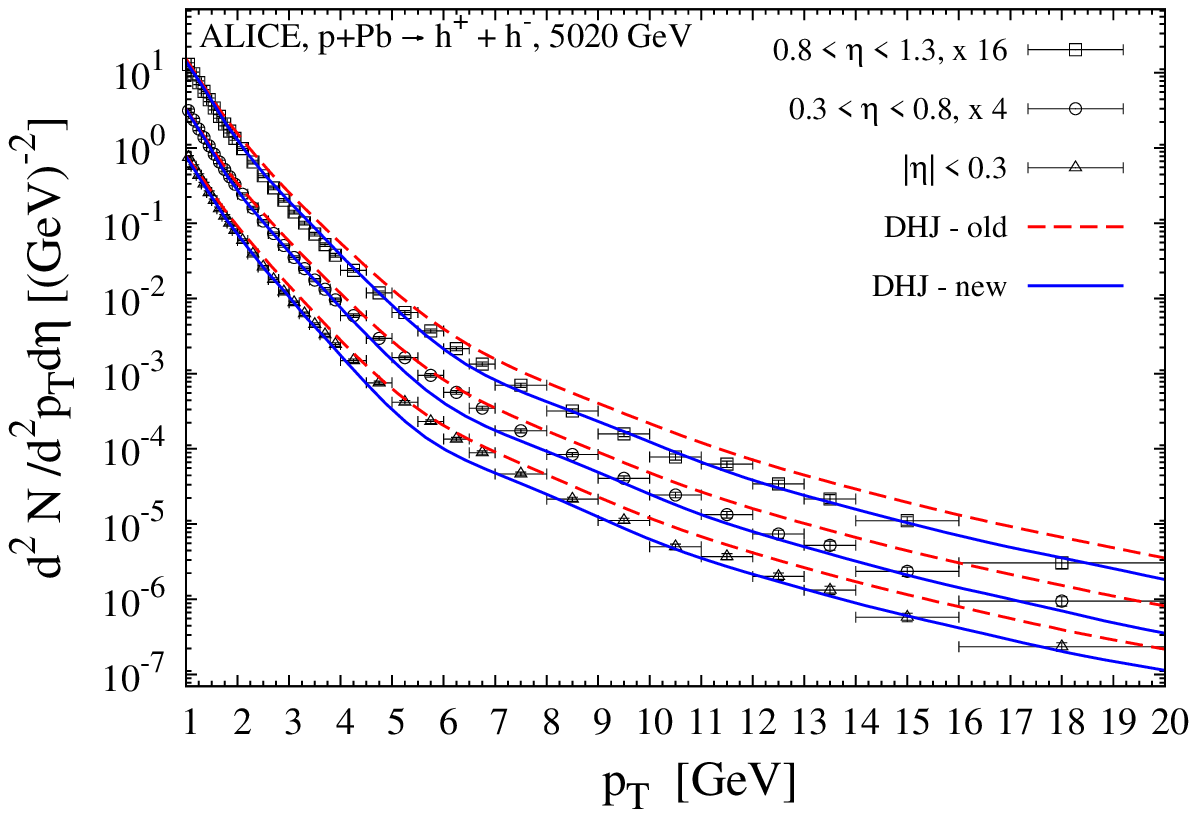}}
\end{center}
\vskip -0.80cm
\caption{(Color online) Comparison between the (a) BUW and (b) DHJ predictions for the transverse 
momentum $p_{T}$-spectra of charged particles produced in $pPb$ collisions and the  ALICE data \cite{alipPb}. 
For the new version of the BUW model we assume $K = 3.7$ for all pseudorapidity bins and for the 
new DHJ model $K = 3.0$, $3.0$ and $3.7$ for $\langle \eta \rangle = 0$, $0.55$ and  $1.05$, respectively. }
\label{fig:1}
\end{figure}

\section{Results and discussion}
\label{section:results}

In what follows we will present our results for the average transverse momentum 
$\langle p_T \rangle$ defined by
\begin{eqnarray}
\langle p_T \rangle = \frac{\int d^2p_T \,p_T \, {dN_h \over dy d^2p_T}}{\int d^2p_T  \, {dN_h \over dy d^2p_T}} \,\,
\end{eqnarray}
which is rapidity and energy dependent, i.e. $\langle p_T \rangle = \langle p_T (y,\sqrt{s})\rangle$. Moreover, 
it depends of the lower limit of the  integrations over the transverse momentum ($p_{T,min}$).  In order to 
obtain realistic predictions for LHC energies it is fundamental to use as input in the calculations a model 
which describes the experimental data on the $p_T$ -- spectra of produced particles. Consequently, as a first 
step we will initially compare the DHJ and BUW predictions with the recent LHC data. In Fig. \ref{fig:1} we 
present a comparison of these predictions using the original parameters, denoted  ``DHJ old'' and  
``BUW old'' in the figures, with the LHC data on the $p_T$ -- spectra of  charged particles 
in $p Pb$ collisions at $\sqrt{s} = 5.02$ TeV and different rapidities \cite{alipPb}. We  use in what follows 
the CTEQ5L parton distribution functions \cite{cteq6}   and  the KKP fragmentation functions \cite{kkp}, with 
the hadron mass being chosen to be the mean value of the pion, kaon and proton masses. Moreover, we  compute 
Eq. (\ref{hybel}) using the central values  of $\eta$ in the  pseudorapidity ranges used in  the experiment and 
choose  $A \equiv A_{\rm min.\, bias} = 20 \, (18.5)$ for $pPb$ $(dAu)$ collisions. We find that these models are 
not able to describe the ALICE data \cite{alipPb} at large transverse momentum with their original parameters. 
The natural next step is to check if a new fit of the free parameters of these models can improve 
the description of the experimental data. As one of the goals of our paper is to check if the average transverse 
momentum can be used to discriminate between the CGC and hydrodynamic descriptions of high multiplicity events 
observed in $pPb$ collisions at LHC, our strategy will be the following:  to determine the free parameters 
of the BUW and the DHJ dipole scattering amplitudes by fitting the $p_{T}$ spectra of charged particles measured 
in $pPb$ collisions at $\sqrt{s} = 5020$ GeV and then  compare the new models with the experimental data on $pp$ 
collisions at  other energies and rapidities. Moreover, differently from the authors of  Refs. \cite{dhj,buw}, 
who  have assumed that $\gamma_s \approx 0.63$, which is the value obtained from the leading order BFKL kernel, 
we will consider $\gamma_s$ as a free parameter. The resulting fits are shown in Figs. \ref{fig:1} (a) and (b)  
for the following parameters: $a = 2.0$, $b = 125 $ and $\gamma_{s} = 0.74$ for the  BUW model and 
$d = 1.0$ and $\gamma_{s} = 0.7$ for the DHJ model. The data are better described if we assume larger 
values of  $\gamma_s \geq 0.7$, which is consistent with the results obtained using the renormalization group improved 
BFKL kernels at next-to-leading order and fixed running coupling \cite{salam}.  As it can be seen, with these parameter 
sets our curves agree well with the experimental data. In the range $4 < p_T < 7$ GeV the 
DHJ curves show an ``edgy'' behaviour which is a reminiscence of the numerical Fourier transform. This is not a big 
effect and can be considered as part of the theoretical error in our calculations. It is important to emphasize 
that $\langle p_T \rangle$  is only marginally affected by these small oscillations (see below) and the qualitative 
fits presented here for both models considered are sufficient to get a realistic prediction for this observable 
since it is dominated by the low $p_{T}$ region.


\begin{figure}[t]
\begin{center}
\subfigure[ ]{
\includegraphics[width=0.45\textwidth]{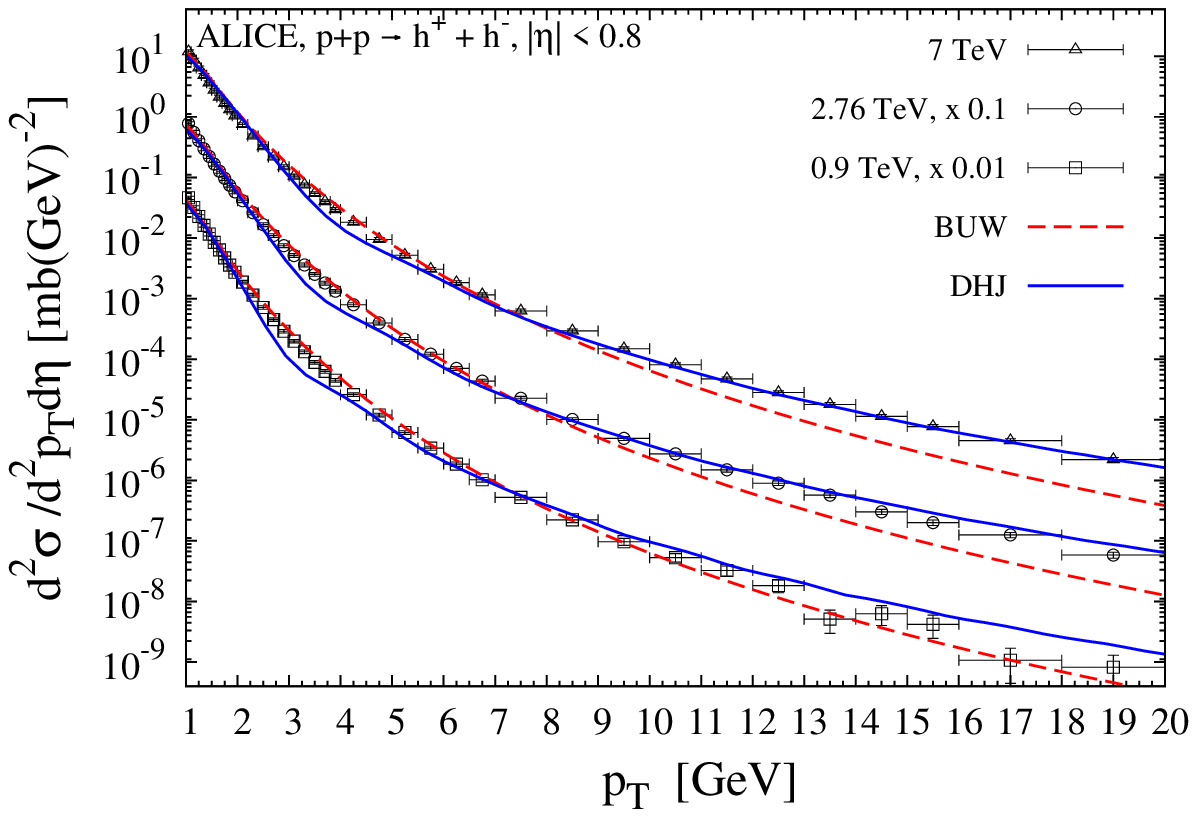}}
\subfigure[ ]{
\includegraphics[width=0.45\textwidth]{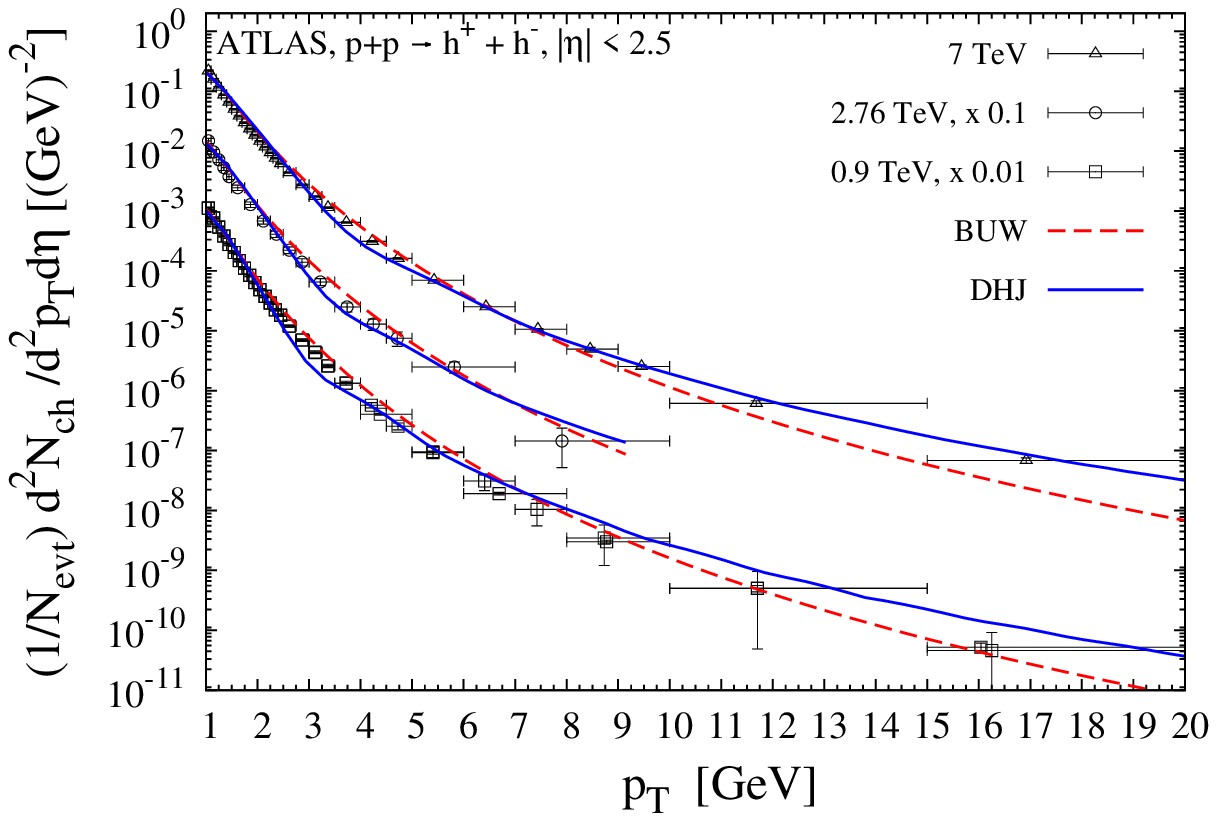}}
\subfigure[ ]{
\includegraphics[width=0.45\textwidth]{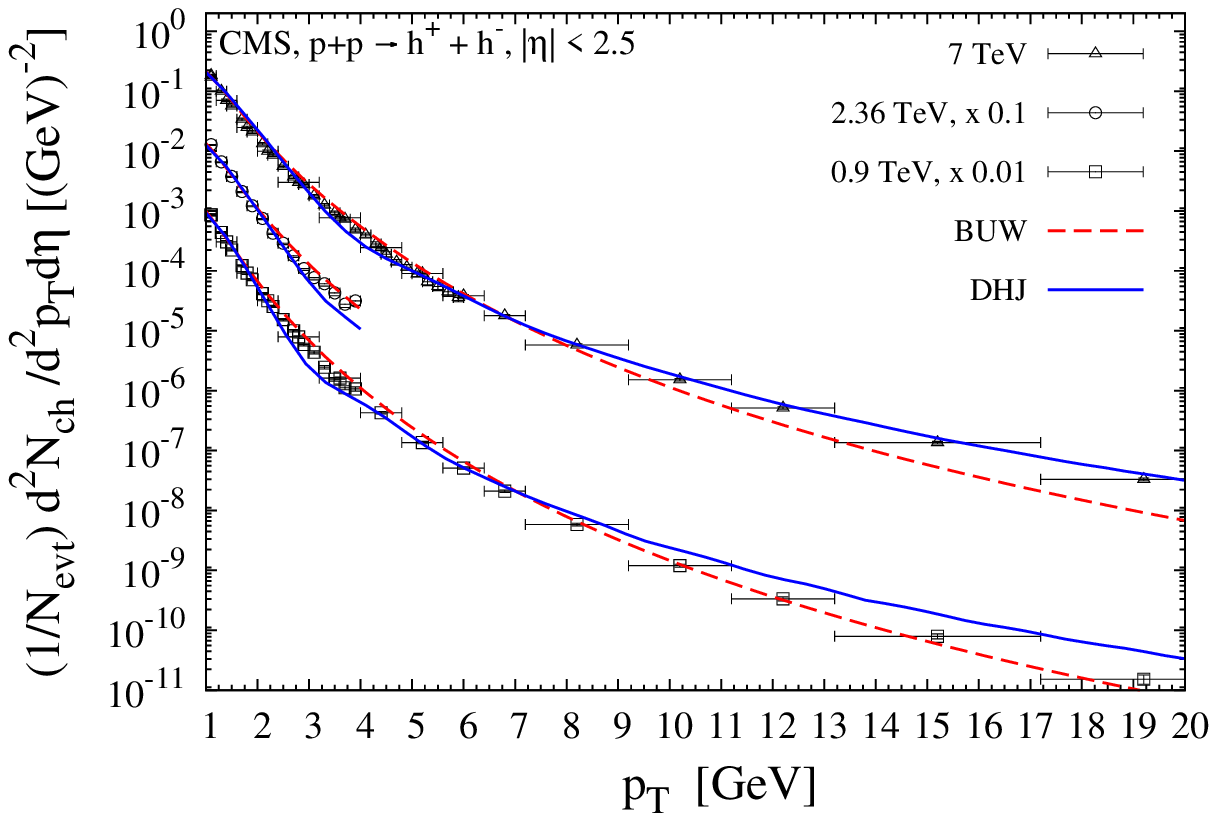}}
\subfigure[ ]{
\includegraphics[width=0.45\textwidth]{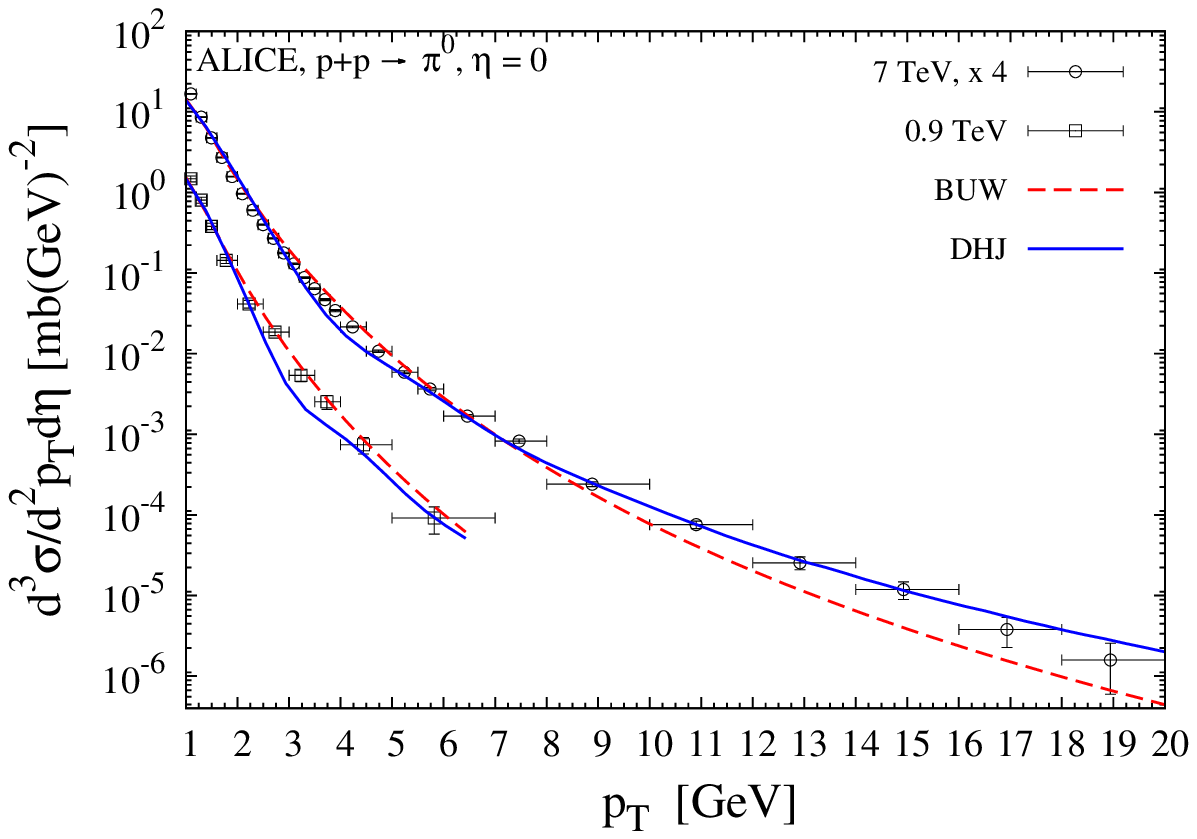}}
\end{center}
\vskip -0.80cm
\caption{(Color online) Predictions of the DHJ and BUW models for the $p_{T}$ spectra of  charged particles   in $pp$ collisions. 
(a) Comparison with the   ALICE data \cite{ali-pp}. The corresponding  $K$ factors are the following: $K = 2.47$ ($K = 2.3$), $2.07$ ($1.85$) 
and $1.77$ ($1.6$) for the BUW (DHJ) model for $\sqrt{s} = 0.9$, $2.76$ and $7$ TeV. (b) Comparison with the ATLAS data \cite{atl-pp}. 
In this case we have assumed  $K = 3.3$,  $2.5$ and $2.3$, respectively, for both models. (c) Comparison with the CMS  data \cite{cms-pp}. 
In this case we assume for both models $K = 3.0$, $2.5$ and $2.3$ for $\sqrt{s} = 0.9$, $2.36$ and $7$ TeV. (d) Predictions of the DHJ and 
BUW models for the $p_{T}$ spectra of  neutral pions. Comparison with the ALICE data \cite{alipi} for $\sqrt{s} = 0.9$ and $7$ TeV. 
For both models we have $K = 1.2$ and 2.0.  }
\label{fig:3456}
\end{figure}

\begin{figure}[t]
\begin{center}
\subfigure[ ]{
\includegraphics[width=0.45\textwidth]{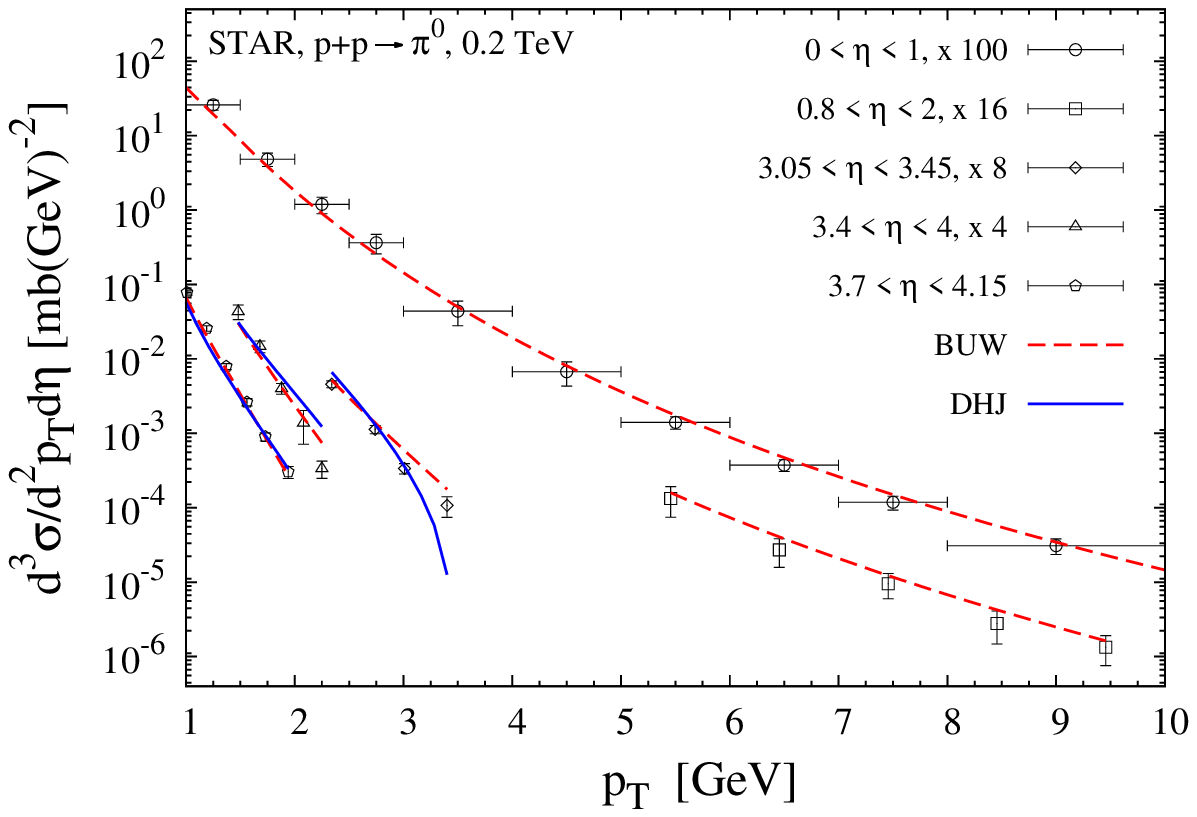}}
\subfigure[ ]{
\includegraphics[width=0.45\textwidth]{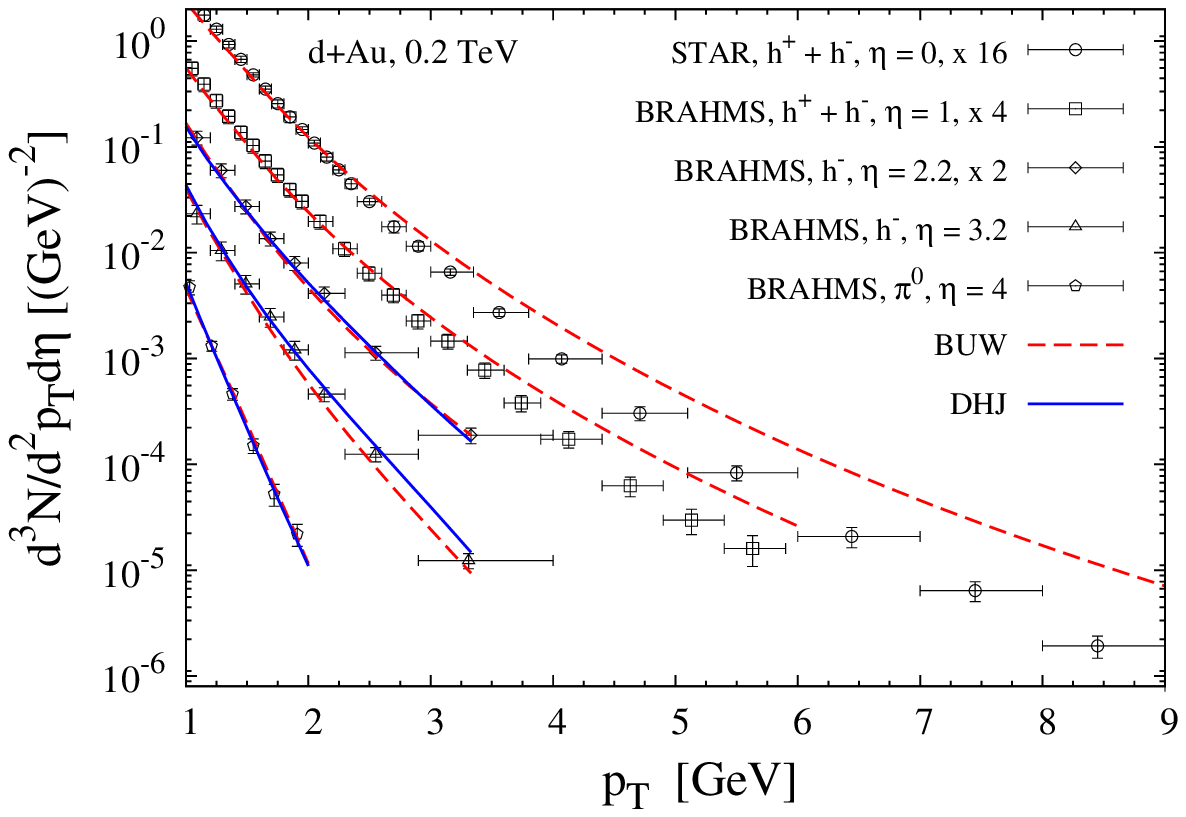}}
\subfigure[ ]{
\includegraphics[width=0.45\textwidth]{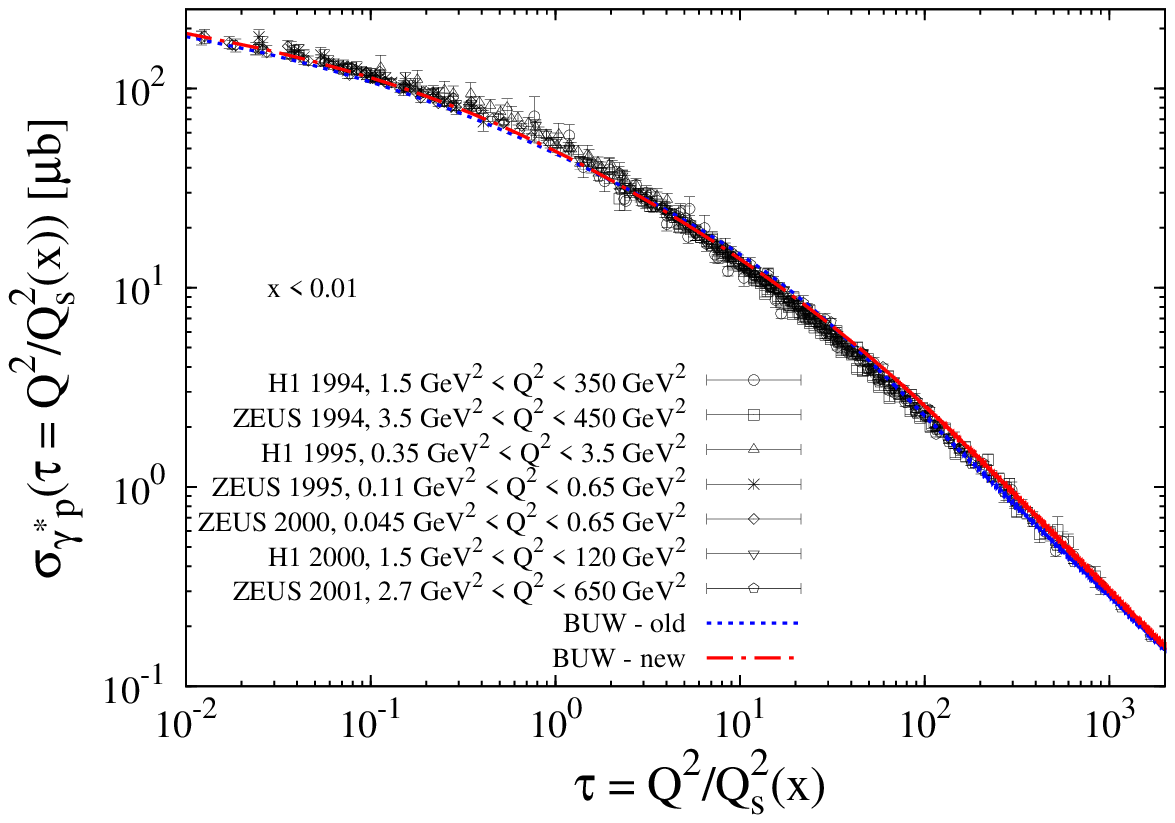}}
\end{center}
\vskip -0.70cm
\caption{ (Color online) (a) Predictions of the DHJ and BUW models for the $p_{T}$ spectra of  neutral pions   in $pp$  collisions. 
For the BUW model $K = 1.5$ for $\langle \eta \rangle = 0.5$ and $K = 1.2$ for 
$\langle \eta \rangle = 1.4, 3.25, 3.7, 3.925$.  For the DHJ model $K = 1.2$ for 
$\langle \eta \rangle = 3.25, 3.7, 3.925$. The experimental data are from \cite{starpi}; 
(b) Predictions of the DHJ and BUW models for the $p_{T}$ spectra of  hadron production   in $dAu$  collisions. 
For the BUW model we have $K = 2.9$, $2.5$, $2.0$, $1.0$ and  $1.0$ for $\eta = 0$, $1$, 
$2.2$, $3.2$ and $4$ respectively. For the DHJ model we have $K = 2.5$, $2.4$, $1.5$ for  $\eta = 2.2$, $3.2$  and $4$ respectively. 
The experimental data are from \cite{dAu}. (c)
Comparison between the BUW predictions and the $ep$ HERA data for the total $\gamma^* p$ cross section \citep{hera-d}.}
\label{fig:2}
\end{figure}

Having fixed the new  parameters of the BUW and DHJ models using the experimental data on hadron production 
in $pPb$ collisions, we now  compare their predictions with the recent LHC data on  $p_{T}$ spectra of charged 
particles and neutral pions measured in $pp$ collisions at different energies and distinct rapidity ranges. 
The only free parameter in our predictions is the $K$ -- factor, which can be energy and rapidity dependent. 
In what follows we will fix this parameter in order to describe the experimental data at lower $p_T$. In 
Fig. \ref{fig:3456} we present our results. We observe that both models  describe quite well the experimental data 
for small $p_T$, with the BUW predictions becoming  worse at higher $p_{T}$ with  increasing  center - of - mass 
energy. In contrast,  the DHJ model also describes quite well data of larger $p_T$, which can be associated to the 
contribution of the geometric scaling violations taken into account in this model. As a final check, let us compare 
the predictions of these new versions of the phenomenological models with the RHIC data on hadron production 
in $pp$ and $dAu$ collisions in the central and forward rapidity regions. In Figs. \ref{fig:2} (a) and (b) we 
present our predictions. We find that both models describe well the experimental data at forward rapidities. On 
the other hand, at central rapidities, the BUW describes well the $pp$ data for $p_T \le 10$ GeV, but fails 
for $p_T \ge 3$ GeV in the case of $dAu$ collisions. In contrast, the results of the DHJ model are not shown for 
these rapidities since they are highly affected by oscilations for $p_{T} \gtrsim 5$ GeV. The failure of the 
description at central rapidities at RHIC is not surprising since the energy is not very large and the formalism 
used here is suited to the study of the  forward region where the small-$x$ component of the target wave function is 
accessed. Finally, in Fig. \ref{fig:2} (c) we demonstrate that the new version of the BUW model satisfies the property 
of the geometric scaling and also is able to describe the $ep$ HERA data for the total $\gamma^* p$ cross section 
in a large range of photon virtualities.

\begin{figure}[t]
\begin{center}
\subfigure[ ]{
\includegraphics[width=0.45\textwidth]{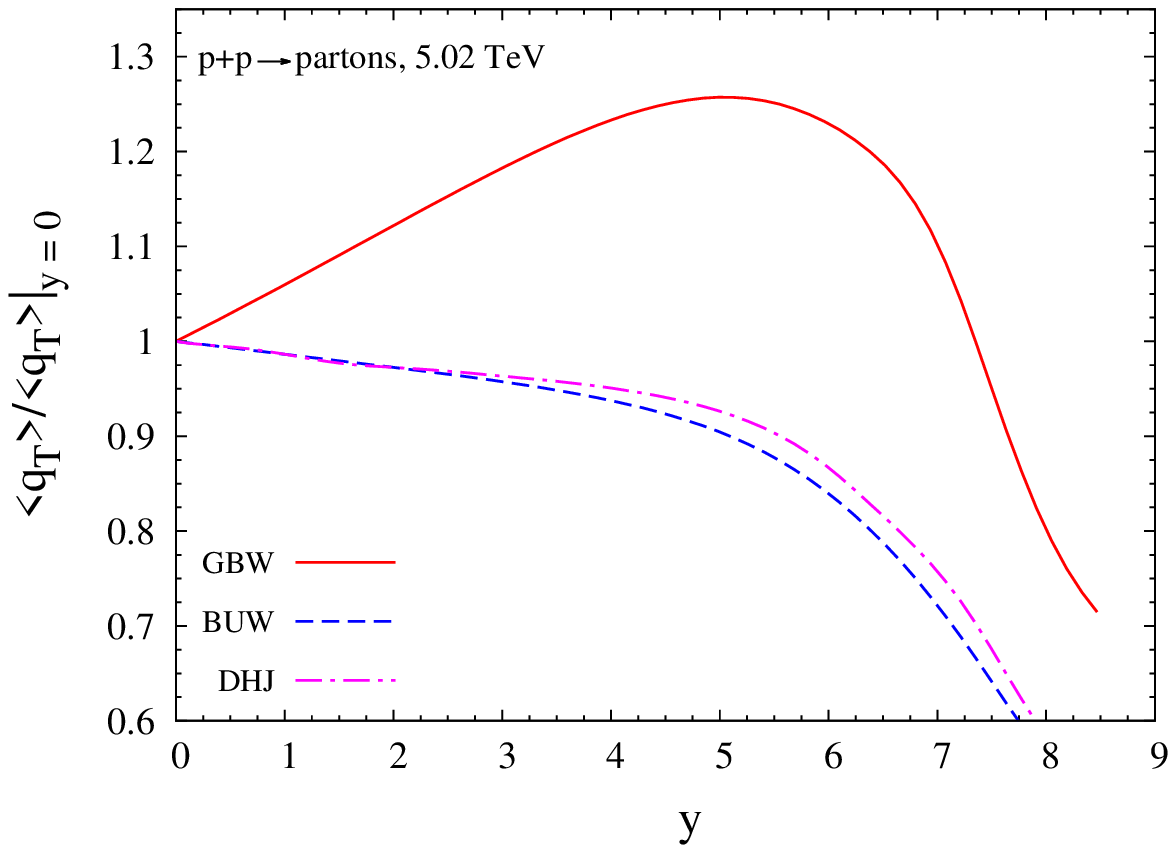}}
\subfigure[ ]{
\includegraphics[width=0.45\textwidth]{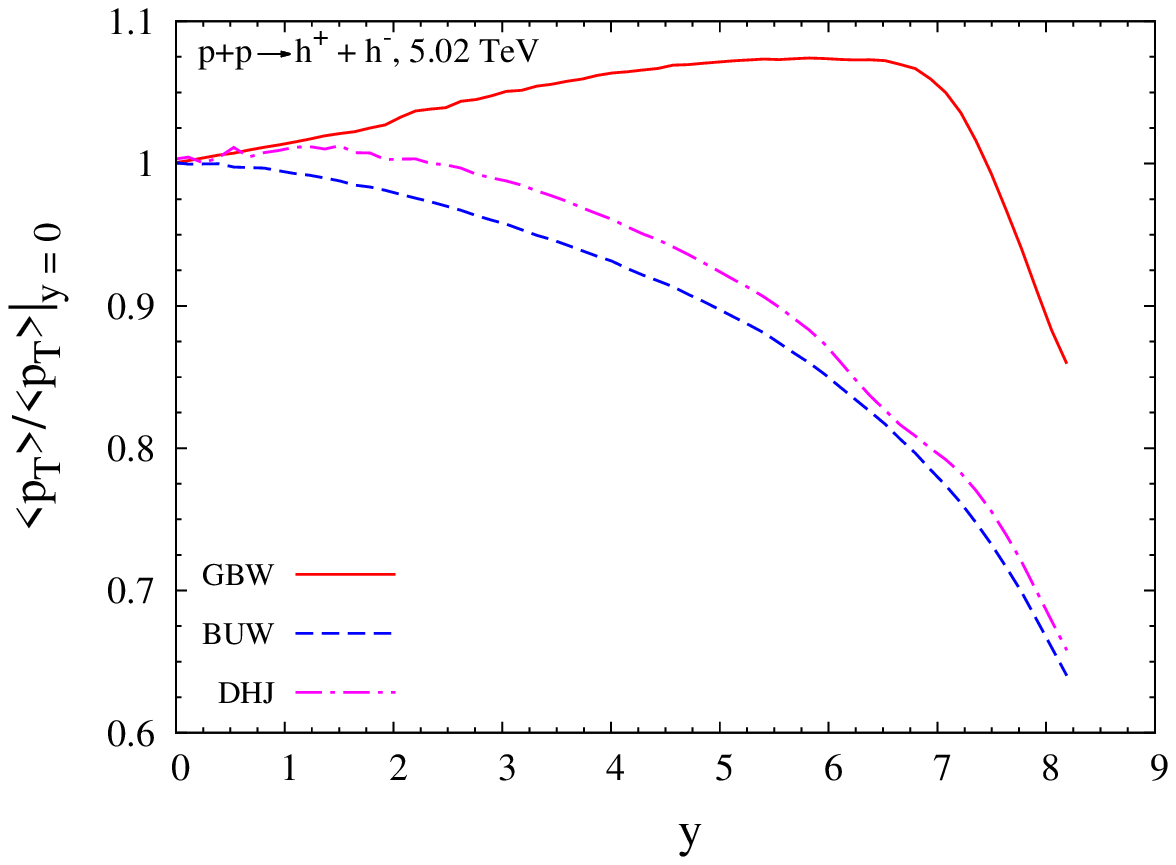}}
\subfigure[ ]{
\includegraphics[width=0.45\textwidth]{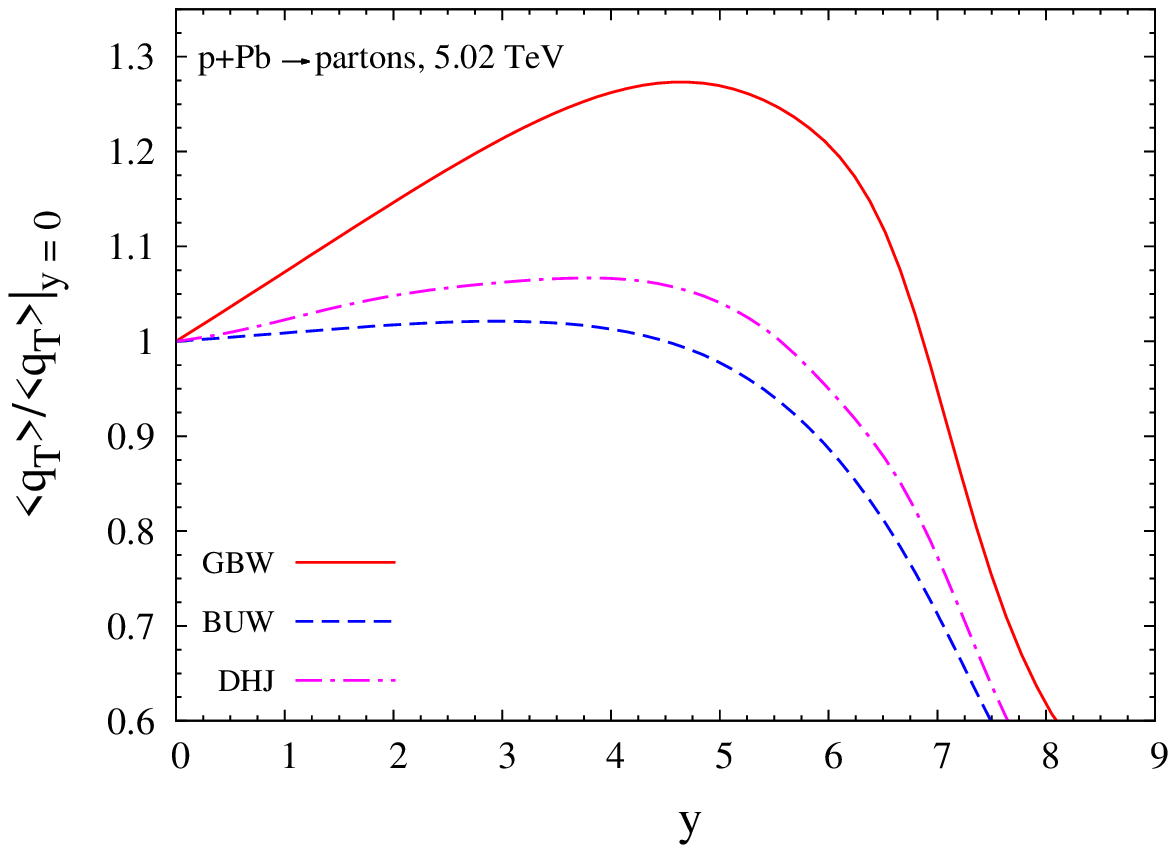}}
\subfigure[ ]{
\includegraphics[width=0.45\textwidth]{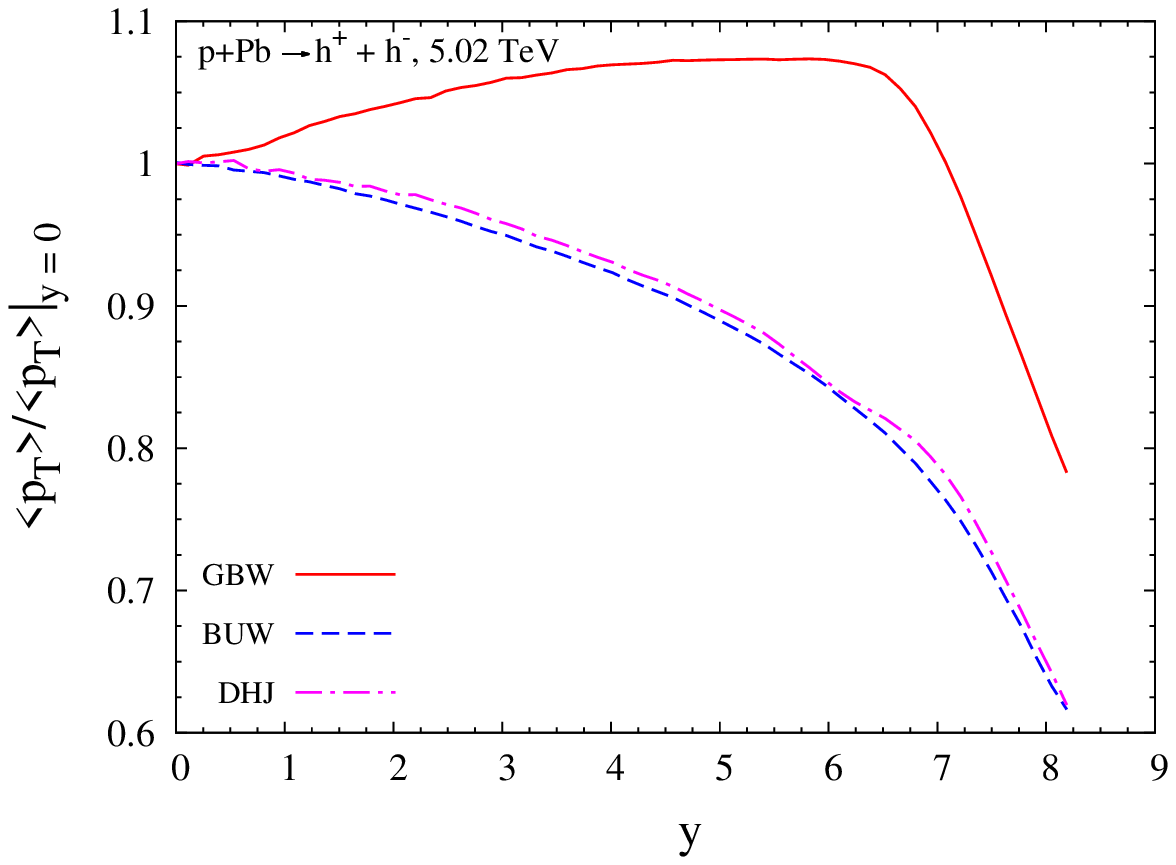}}
\end{center}

\vskip-0.7cm
\caption{(Color online) Dependence of the ratio $\langle p_{T}(y, \sqrt{s})\rangle / \langle p_{T}(0, \sqrt{s})\rangle$ in the model used for the 
forward scattering amplitude in $pp$  and $pPb$ collisions. In  panels (a) and (c)  parton fragmentation is disregarded, while in (b) and (d) fragmentation is included. }
\label{fig:ptmedpar}
\end{figure}

\begin{figure}[t]
\begin{center}
\subfigure[ ]{
\includegraphics[width=0.45\textwidth]{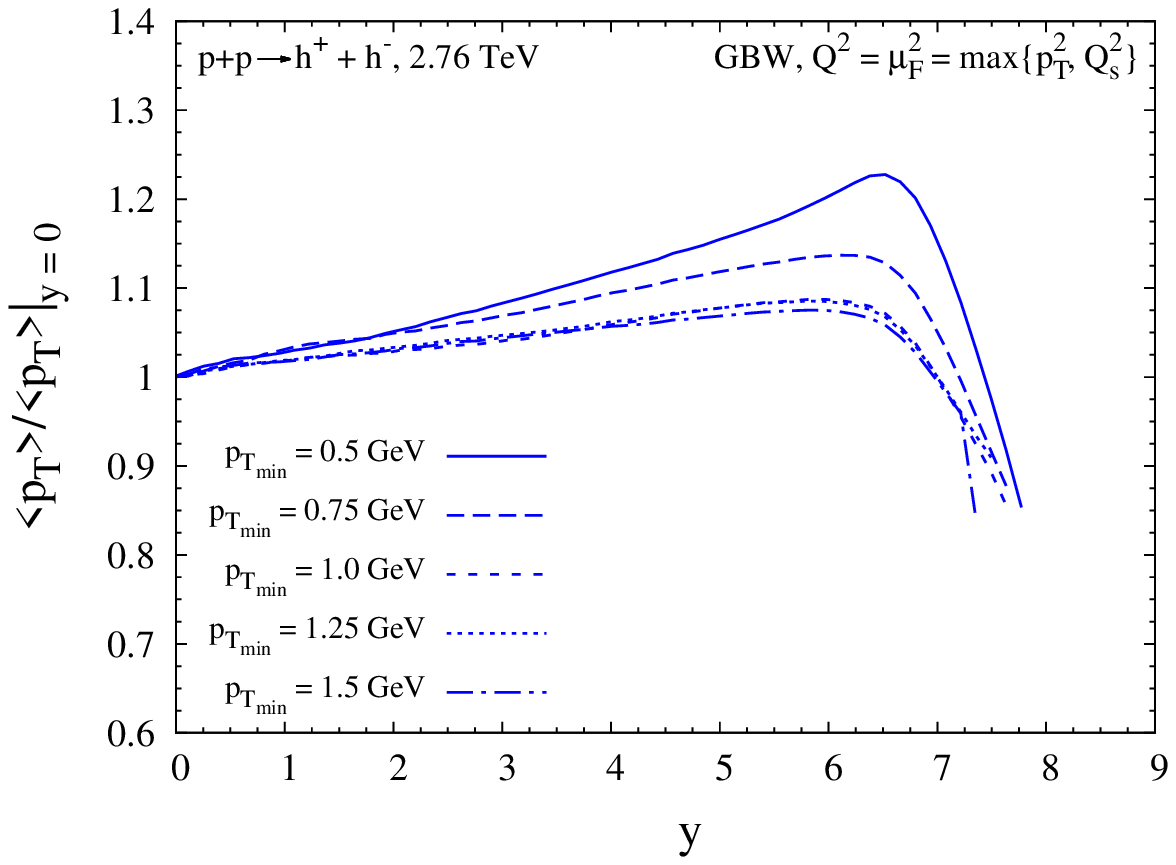}}
\subfigure[ ]{
\includegraphics[width=0.45\textwidth]{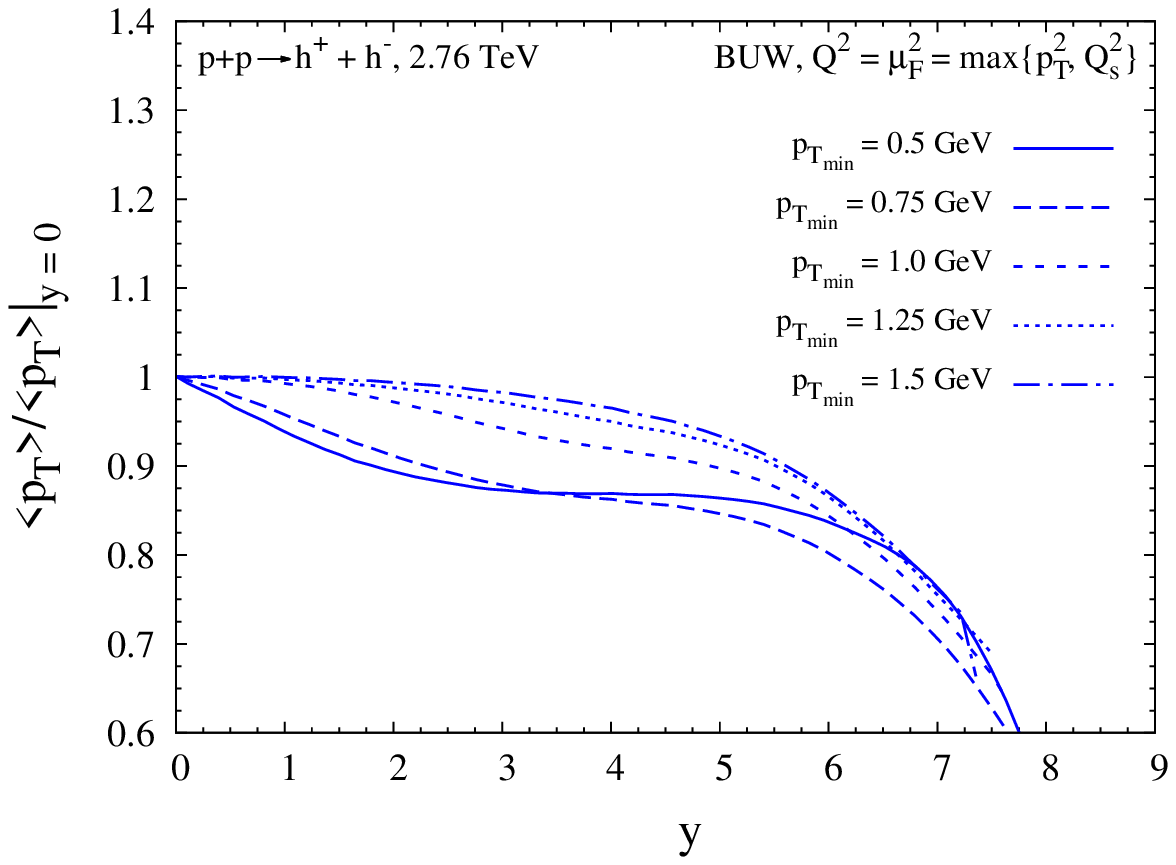}}
\end{center}
\vskip-0.7cm
\caption{(Color online) Dependence of the ratio $\langle p_{T}(y, \sqrt{s})\rangle / \langle p_{T}(0, \sqrt{s})\rangle$ 
in the minimum transverse momentum $p_{T,min}$ considering (a) the GBW and (b) the BUW model for the forward scattering amplitude. }
\label{fig:ptmin}
\end{figure}

The results presented in Figs. \ref{fig:1} -- \ref{fig:2} make us confident to obtain realistic predictions 
for the average transverse momentum.  In what follows we will study the energy and rapidity dependencies of the ratio
\begin{equation}
R = \frac{\langle p_{T}(y, \sqrt{s})\rangle}{\langle p_{T}(0, \sqrt{s})\rangle}
\label{rat}
\end{equation}
where the denominator represents the average transverse momentum at zero rapidity. The motivation to 
estimate this ratio is the reduction of the  uncertainties related to the fragmentation functions as well 
as in the choice of the minimum transverse momentum present in the calculation of $\langle p_{T} \rangle$. 
Initially, let us analyse the dependence of our predictions on the model used to describe the forward 
scattering amplitudes ${\cal{N}_{A,F}}(x,r)$  and the impact of the inclusion of  parton fragmentation. In Fig. \ref{fig:ptmedpar} we compare the predictions of the 
BUW and DHJ models with those  from the GBW model \cite{gbw}, obtained  assuming $p_{T,min} = 1$ GeV. It is 
important to emphasize that the GBW model is not able to describe the experimental data on  hadron 
production in hadronic collisions, since it predicts that ${\cal{N}_{A,F}}(x,r)$  decreases exponentially 
at large transverse momentum. However, as this model is usually considered to obtain analytical results 
for several observables, we would like to verify if its predictions for  $\langle p_{T} \rangle$ are realistic. 
In Fig. \ref{fig:ptmedpar} (a) and (c) we present our predictions disregarding  parton fragmentation, while in  panels (b) and (d) fragmentation is included. 
It is important to emphasize that our results for the GBW model  without fragmentation, obtained using the hybrid formalism are similar to those obtained in 
Ref. \cite{bbs} with the $k_T$ - factorization approach.
We can see that the DHJ and BUW predictions are similar (to each other) and differ significantly from the GBW one.
While the GBW model predicts a growth of the ratio for $y \le 6$, the BUW and DHJ models 
predict that this  ratio is almost constant or decreases  with  rapidity. The inclusion of  parton fragmentation modifies the rapidity dependence, 
implying a smaller growth of the GBW prediction. In the case of the DHJ and BUW predictions, the inclusion of  fragmentation implies that the fall of the ratio 
begins at smaller rapidities. Our results demonstrate that the inclusion of  fragmentation has an important impact on the behavior of $\langle p_{T} \rangle$. 
However, the main difference between our predictions and those presented in Ref. \cite{bbs} comes from the model used to describe the QCD dynamics at high energies. 
This distinct behavior is present for $pp$ and $pPb$ collisions, with the behavior of the ratio at very large rapidities being determined by 
kinematical constraints associated to the limited phase space. 
These results were obtained considering $p_{T,min} = 1$ GeV. In Fig. \ref{fig:ptmin} we analyse the dependence of our results on this arbitrary cut off 
in the transverse momentum. For this calculation we have compared the value of the saturation scale for a 
given transverse momentum and rapidity with the corresponding value of $p_T$ and assumed that the factorization 
scale is given by the harder scale. This basic assumption has been used in Ref. \cite{mario} in order to 
extend the hybrid formalism to hadron production at very small - $p_T$, obtaining a very good description 
of the LHCf data. However, it is important to emphasize that we have checked that similar results are obtained 
if we freeze the factorization scale at the minimum value of $Q^2$ allowed in the parton distributions and 
fragmentation functions when smaller values of $p_T$ are probed in the calculation.   The results shown in 
Fig. \ref{fig:ptmin} indicate that the behaviour of the ratio with the rapidity is not strongly modified by 
the choice of $p_{T,min}$. Consequently,  we will consider $p_{T,min} = 1$ GeV in what follows.

\begin{figure}[t]
\begin{center}
\subfigure[ ]{
\includegraphics[width=0.45\textwidth]{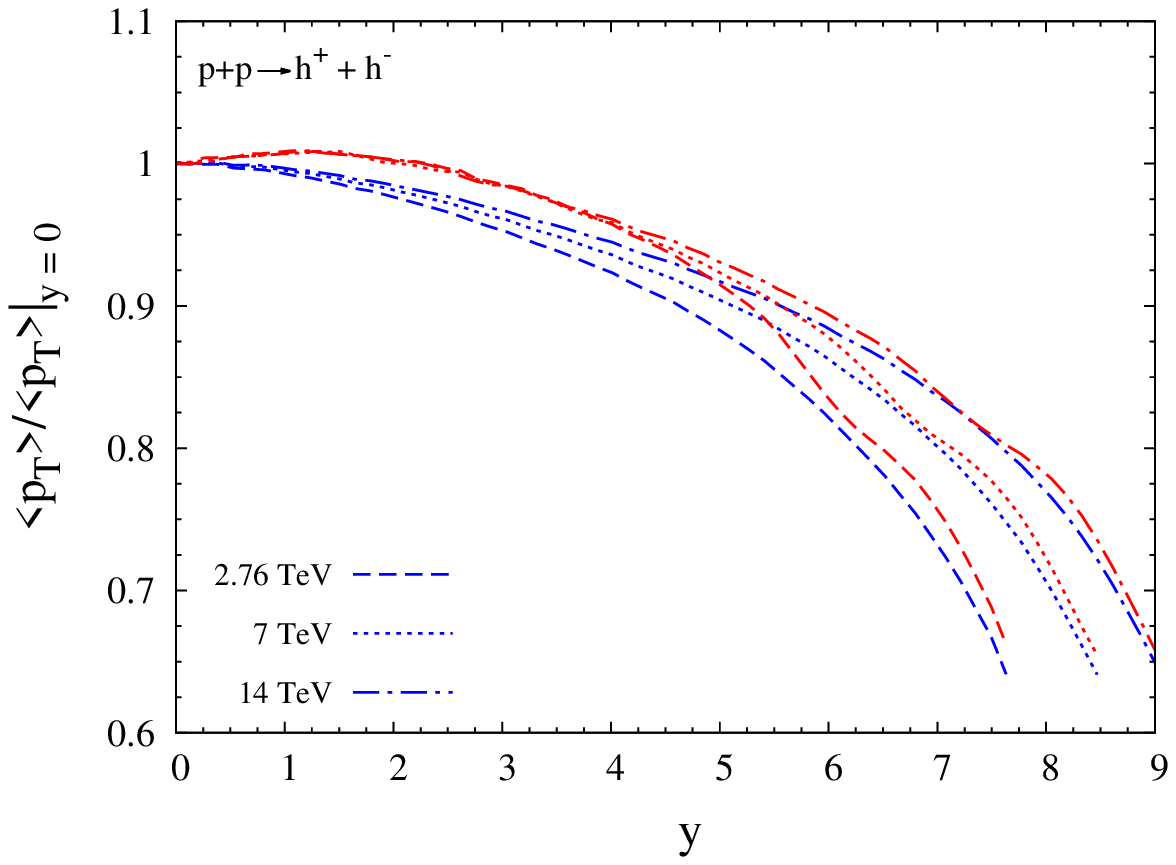}}
\subfigure[ ]{
\includegraphics[width=0.45\textwidth]{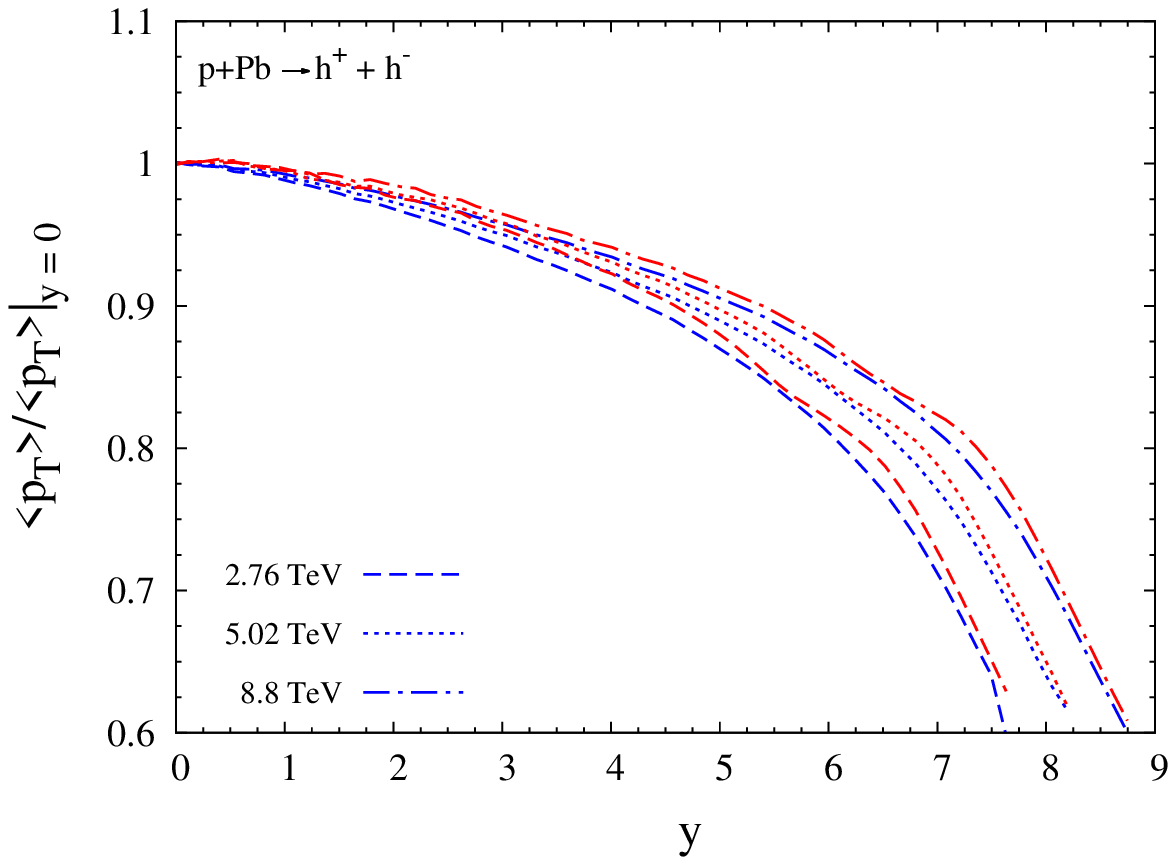}}
\end{center}
\vskip-0.65cm
\caption{(Color online) Rapidity dependence of the ratio $R = \langle p_{T}(y, 
\sqrt{s})\rangle / \langle p_{T}(0, \sqrt{s})\rangle$ in (a) $pp$ and (b) $pPb$ 
collisions for different energies. The BUW (DHJ) predictions are represented by 
blue (red) lines.}
\label{fig:ptmed}
\end{figure}

\begin{figure}[t]
\begin{center}
\subfigure[ ]{
\includegraphics[width=0.45\textwidth]{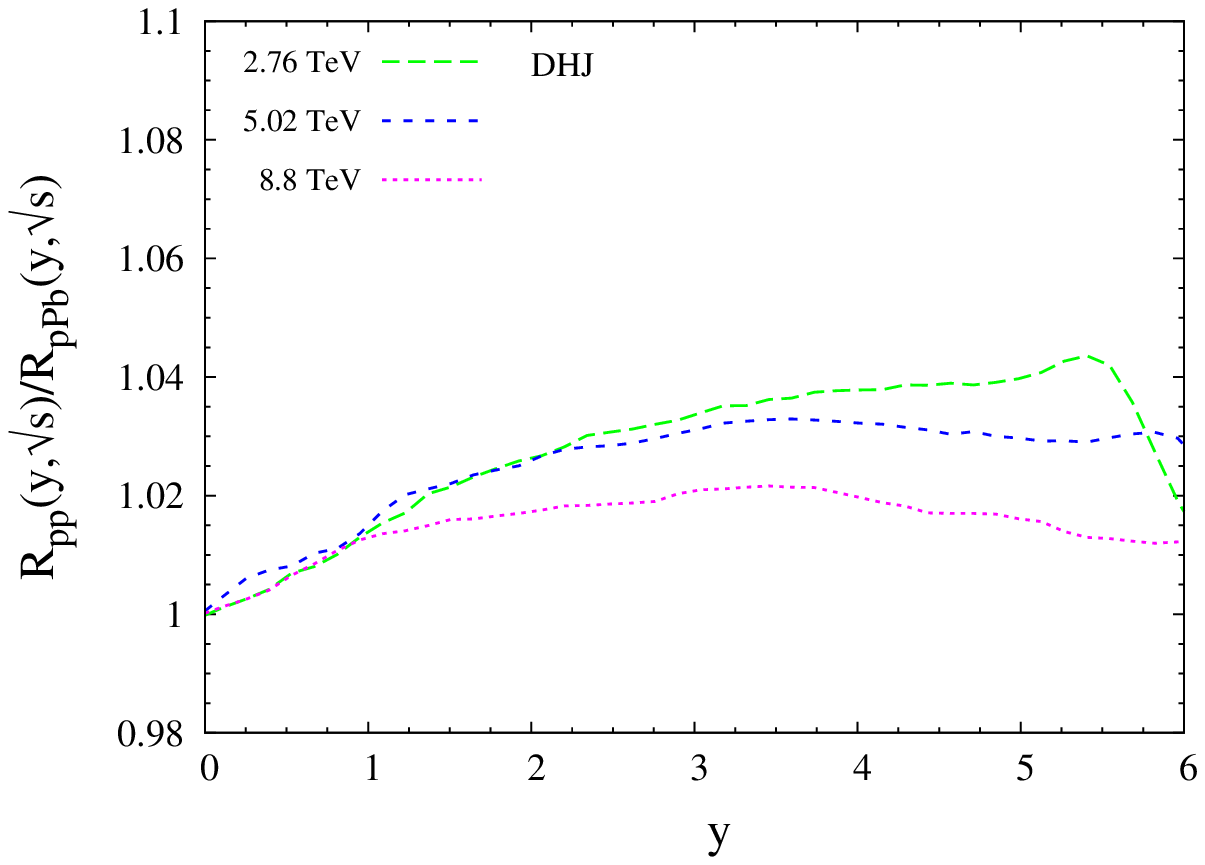}}
\subfigure[ ]{
\includegraphics[width=0.45\textwidth]{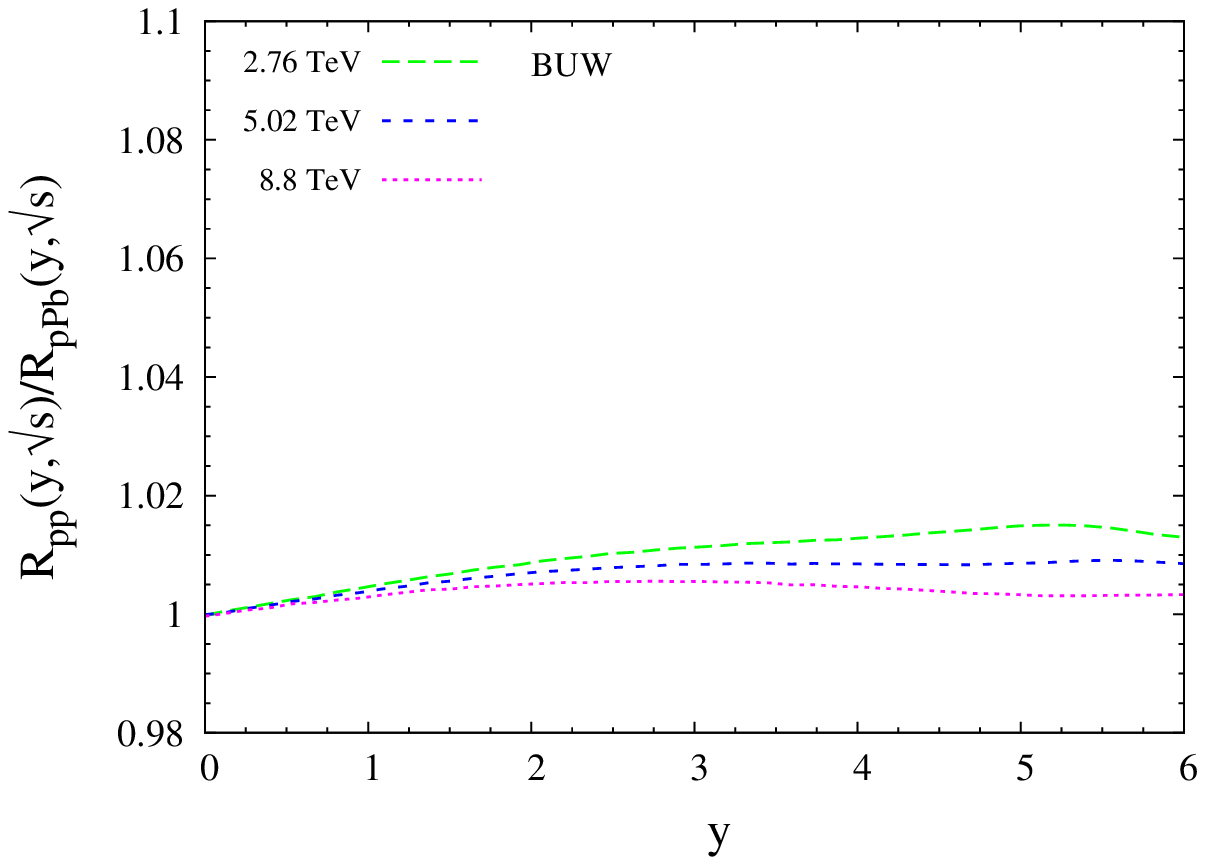}}
\end{center}
\vskip-0.65cm
\caption{(Color online) Rapidity dependence of the ratio between the predictions for the $R = \langle p_{T}(y, \sqrt{s})\rangle / \langle p_{T}(0, \sqrt{s})\rangle$ in $pp$ collisions and those for $pPb$ collisions considering (a) the DHJ and (b) the BUW models.}
\label{fig:ratiopts}
\end{figure}

In Fig.  \ref{fig:ptmed} we present the behaviour the ratio 
$\langle p_{T}(y, \sqrt{s})\rangle / \langle p_{T}(0, \sqrt{s})\rangle$ for $pp$ and $pPb$ collisions 
considering different center of mass  energies. We find that the predictions of the DHJ (red lines) 
and BUW (blue lines) are similar, with the DHJ being slightly larger than the BUW, and that the ratio 
increases with energy. Moreover, we oberve that for a fixed energy the ratio is larger for $pp$ in 
comparison to $pPb$ collisions, as demonstrated in Fig. \ref{fig:ratiopts} where  we present our 
results for the ratio between the predictions for 
$R = \langle p_{T}(y, \sqrt{s})\rangle / \langle p_{T}(0, \sqrt{s})\rangle$ in $pp$ collisions and those 
obtained for $pPb$ collisions. Our results indicate that at very large energies the predictions for $R$ 
in $pp$ and $pPb$ collisions become identical. These predictions are an important test of the hybrid 
factorization and the CGC formalism. We believe that the analysis of the ratio $R$ in $pp$ and $pPb$ 
collisions can be useful to   probe the QCD dynamics at forward rapidities. Finally, the  results from 
Fig.  \ref{fig:ptmed} indicate that the ratio decreases with the rapidity in $pPb$ collisions for the 
energies probed by LHC, presenting a behaviour similar to that obtained using a hydrodynamical approach, 
which implies that in principle this observable cannot be used to discriminate the CGC and hydrodynamical 
approaches for the description of the high multiplicity events. This conclusion is opposite to that obtained 
in Ref. \cite{bbs}. This difference comes from several facts. First, the CGC results 
in  Ref. \cite{bbs}  were obtained using an analytical approximation for a particular  unintegrated gluon 
distribution that does not describe (even at a qualitative level) the experimentally measured  $p_T$-spectra. 
Second,  the calculation presented in \cite{bbs}  does not include the important contribution of the fragmentation 
processes to the average transverse momentum. Finally, kinematical constraints associated with phase space restrictions 
at large rapidities were not included in \cite{bbs} and, even in a partonic scenario, they play an important role at very large rapidities. 
In contrast, in our analysis we have calculated the ratio $R$ using  two 
different models for the forward scattering amplitude that are able to 
describe the current experimental data on charged hadron and pion $p_T$ spectra measured in $pp$ and $pPb$ 
collisions at LHC.  We have included the effects of parton fragmentation and phase space restrictions. It is important to 
emphasize that although we have used the hybrid formalism instead of the $k_T$ - factorization approach, we have verified that both approaches 
imply a similar behavior for the ratio $R$ when the GBW model is used as input and the parton fragmentation is not taken into account. 
Our results demonstrate that the main difference comes from the treatment of the QCD dynamics at high energies.


\section{Conclusions}
\label{section:conc}

In this paper we considered the hybrid formalism to study the behaviour of 
the average $p_{T}$ with the rapidity in $pp$ and $pPb$ collisions at  several 
energies in the CGC picture of high energy collisions. In order to obtain realistic 
predictions we have updated previous phenomenological models for the forward 
scattering amplitude,  one with and other without geometric scaling violations. 
After constraining  their parameters with the most recent data on the $p_{T}$ spectra 
of  charged particles, measured in $pPb$ collisions at the LHC, we demonstrated 
that they are able to describe the recent $pp$ data on the charged hadron and 
pion $p_T$ spectra measured at LHC in the kinematical range of $p_T \le 20$ GeV. Comparison 
of their predictions with the HERA and RHIC  data were also presented. Using 
these models as input, we have calculated  the average transverse momentum 
$\langle p_{T}(y, \sqrt{s})\rangle$ in $pp$ and $pPb$ collisions, and estimated the 
energy and rapidity dependencies of the 
$R = \frac{\langle p_{T}(y, \sqrt{s})\rangle}{\langle p_{T}(0, \sqrt{s})\rangle}$, which is 
an observable  that can be analysed experimentally. We demonstrated that this ratio increases 
with the energy for a fixed rapidity and decreases with the rapidity for a fixed energy, with 
a behaviour similar to that predicted in hydrodynamical approaches for high multiplicity events. 
Our results indicated that this decreasing comes from the treatment of the QCD dynamics at high energies and the inclusion of the  fragmentation 
process and  kinematical constraints associated to the phase space restrictions at very large 
rapidities.
Finally, we  demonstrated that these behaviours are  very similar in $pp$ and $pPb$ collisions.

\section*{Acknowledgments}

This work was partially financed by the Brazilian funding agencies CAPES, CNPq, 
FAPESP and FAPERGS. We thank Dani{\"e}l Boer for very instructive discussions.

\end{document}